\documentstyle[supertabular,epsf,rotate]{mn}
\newcommand{\f}{\phantom{2}}
\newcommand{\mc}{\multicolumn}

\newcommand{\ltsimeq}{\raisebox{-0.6ex}{$\,\stackrel 
        {\raisebox{-.2ex}{$\textstyle <$}}{\sim}\,$}} 
\newcommand{\gtsimeq}{\raisebox{-0.6ex}{$\,\stackrel 
        {\raisebox{-.2ex}{$\textstyle >$}}{\sim}\,$}} 

\input epsf.sty

\begin{document}

\title[Submillimetre photometry of high-redshift radio quasars]
{Submillimetre photometry of typical high-redshift radio quasars} 

\author[Rawlings et al.]{
Steve Rawlings$^{1}$\footnotemark, 
Chris J.\ Willott$^{1}$, Gary J.\ Hill$^{2}$, \\
\\
{\LARGE
Elese N.\ Archibald$^{3}$, James S.\ Dunlop$^{4}$ and
David H.\ Hughes$^{5}$}\\
\\
$^{1}$ 
Astrophysics, Department of Physics, Denys Wilkinson Building, Keble Road,
Oxford, OX1 3RH, UK \\ 
$^{2}$ McDonald Observatory, Department of Astronomy, University of Texas at 
Austin, Austin, TX 78712, USA\\ 
$^{3}$ Joint Astronomy Centre, 660 N. A`oh\={o}k\={u}
Place, University Park, Hilo, Hawaii, 9620, USA \\ 
$^{4}$ Institute for
Astronomy, University of Edinburgh, Blackford Hill, Edinburgh, EH9
3HJ, UK\\
$^{5}$ Instituto Nacional de Astrofisica, Optica y Electronica 
(INAOE), Apartado Postal 51 y 216, 72000 Puebla, Pue., Mexico
}

\maketitle

\begin{abstract}
\noindent
We present submillimetre (SCUBA) photometry of a sample of eight high-redshift 
($2.5 \leq z < 3.5$) radio quasars from two redshift surveys: 
the {\sc TexOx}-1000, or TOOT, survey;
and the 7C Redshift Survey (7CRS). Unlike the powerful
high-redshift radio sources observed previously in the submillimetre, these 
radio sources are typical of those dominating the radio luminosity density
of the population. We detect just two of the TOOT/7CRS
targets at 850 $\mu\rm{m}$, and one of these detections is probably due to 
synchrotron emission rather than dust. The population 
represented by the
other six objects is detected in a statistical sense with
their average 850 $\mu\rm{m}$ flux density implying that they 
are similar to low-redshift, far-infrared-luminous quasars undergoing at 
most moderate ($\ltsimeq 200 ~ \rm M_{\odot} ~ yr^{-1}$) starbursts. 
By considering all the SCUBA data available for radio sources,
we conclude that positive correlations
between rest-frame far-infrared luminosity
$L_{\rm FIR}$, 151-MHz luminosity $L_{151}$ and redshift $z$, although
likely to be present, are hard to interpret because of 
subtle selection and classification biases, small number
statistics and uncertainties concerning synchrotron contamination
and $K$-correction. We argue that there is not yet any
compelling evidence for significant differences in 
the submillimetre properties of radio-loud and radio-quiet quasars at
high redshift.
\end{abstract}


\footnotetext{Email: s.rawlings1@physics.ox.ac.uk}

\section{Introduction}

The advent of the Submillimetre Common-User Bolometer Array
(SCUBA; Holland et al.\ 1999) camera on the James Clerk Maxwell
Telescope (JCMT) opened up the field of submillimetre cosmology.
This, and other, instrument/telescope combinations now routinely detect 
dusty star-forming galaxies at high redshifts, a process greatly aided
by the way in which the steep submillimetre spectral index 
compensates for distance-dependent dimming, so that two objects
of similar far-infrared luminosity have similar observed submillimetre
flux density if both lie anywhere in the redshift 
range $1 \ltsimeq z \ltsimeq 10$ (Blain \& Longair 1993).

A major problem in understanding how submillimetre surveys constrain
the history of star formation in the Universe is the difficulty of
measuring redshifts for the sources of the submillimetre emission. 
This problem has not yet been solved, with only small ($< 10$ per cent)
fractions of objects in submillimetre-flux-density-limited samples having 
reliable spectroscopic redshifts (e.g. Blain et al.\ 2002).

An alternative strategy for studying the cosmic star-formation history is
to make targeted submillimetre observations of objects with known 
redshifts (e.g.\ Hughes, Dunlop \& Rawlings 1997). 
Powerful radio sources are important targets, not because 
the amount of star formation associated with them seems likely to be
cosmologically significant, but because of two crucial
properties: (i) they appear to be associated with galaxies, and
also presumably black holes, with a mass distribution with a large mean and a 
small dispersion (e.g.\ Jarvis et al.\ 2001b); 
and (ii) studies of the physics of their extended 
radio structures allows the calculation of the time since their 
jet-producing AGN were first triggered 
(e.g.\ Blundell \& Rawlings 1999).
The first property is potentially very useful for determining
how physical properties derived from submillimetre observations
(e.g.\ dust mass and star-formation rate) depend on galaxy, or 
black hole, mass. The second property can be used to constrain 
models in which starburst and AGN activity are physically related.

The first large study of the submillimetre emission of 
high-redshift radio sources was made by
Archibald et al.\ (2001). They used SCUBA on the JCMT
to make sensitive (to an 850 $\mu$m noise level $\approx 1$ mJy rms) 
observations of 47 radio galaxies covering the redshift
range $1<z<4.5$. The chief result of their study was 
a rise in the detection rates of radio sources 
from $\approx 15$ per cent at $z \ltsimeq 2.5$ to 
$\gtsimeq 75$ per cent at higher redshifts. They interpreted their
results as indicating that the submillimetre luminosity of radio
galaxies, and hence physical quantities like dust mass,
increases systematically with redshift. 

One important limitation of the Archibald et al.\ (2001) study
was that it targeted only radio
galaxies and excluded radio quasars\footnotemark. This was largely a practical 
decision, because it was thought that the bright radio-emitting cores
of quasars might produce troublesome synchrotron contamination.
The decision could be defended on the basis of the simplest
unified scheme (e.g.\ Antonucci 1993) in which no
differences are predicted in the joint distributions of
isotropically-emitted quantities like synchrotron low-frequency-radio and
dust submillimetre emission. A follow-up study of the submillimetre properties of powerful
radio quasars at $z \sim 1.5$ (Willott et al.\ 2002a) revealed a
significant difference between the submillimetre
luminosities of radio-luminosity-matched quasars and galaxies --
the quasars are brighter in the submillimetre by a factor $\geq 2$,
accounting for synchrotron contamination.
Willott et al.\ show that their result can still be accommodated within a
unified scheme provided one allows for systematic changes in the
properties of the obscuring material with time after the
jet-triggering event (e.g.\ Blundell \& Rawlings 1999; Hirst, Jackson 
\& Rawlings 2003) and/or correlations between quasar
luminosity and the probability of viewing a given object as a quasar
(e.g.\ as in the `receding torus' model; Lawrence 1991, Simpson 1998).

A second important limitation of the Archibald et al.\ (2001) study was
that it was confined only to the most radio-luminous sources:
such objects contribute marginally to the
luminosity density of the radio source population and
could plausibly be entirely unrepresentative. 
The main aim of this paper is to establish the submillimetre 
properties of typical high-redshift radio sources. 

\footnotetext{
We will use the term radio quasar to refer to lobe-dominated
steep-spectrum radio sources associated with quasar nuclei.
These are the dominant population of radio-loud quasars in the
low-frequency-selected samples considered in this paper, 
whereas compact flat-spectrum 
quasars can dominate in high-frequency-selected samples.
Unified schemes 
(e.g.\ Antonucci 1993) invoke an optically-thick obscuring torus
which hides the quasar nucleus when the jet axis makes a large angle with the
line-of-sight, and such objects are termed radio galaxies.
}

In Sec.~\ref{sec:sample}
we describe how a sample of radio sources was selected for SCUBA 
follow-up, and how these observations were made. The results of these
observations are presented and discussed. In
Sec.~\ref{sec:toot} we use these results to establish the
submillimetre properties of typical high-$z$ radio sources,
to look for evidence of correlations between far-IR luminosity,
radio luminosity and redshift and
to compare and contrast these results with those obtained
previously by Archibald et al.\ (2001) and Willott et al.\ (2002a).
Some concluding remarks are made in Sec.~\ref{sec:conclusions}.

We use J2000.0 positions throughout.
The convention for all spectral 
indices, $\alpha$, is that flux density $S_{\nu} \propto \nu^{-\alpha}$,
where $\nu$ is the observing frequency. 
We assume throughout a low-density, $\Lambda$-dominated 
Universe in which
$H_{\circ}=70~ {\rm km~s^{-1}Mpc^{-1}}$, $\Omega_{\rm M}=0.3$ and $\Omega_
{\Lambda}=0.7$.

\section{SCUBA observations of high-redshift TOOT and 7CRS radio sources}
\label{sec:sample}

\subsection{Selection of SCUBA targets}
\label{sec:selection}

\begin{figure*}
\begin{center}
\setlength{\unitlength}{1mm}
\begin{picture}(150,120)
\put(-50,-30){\includegraphics{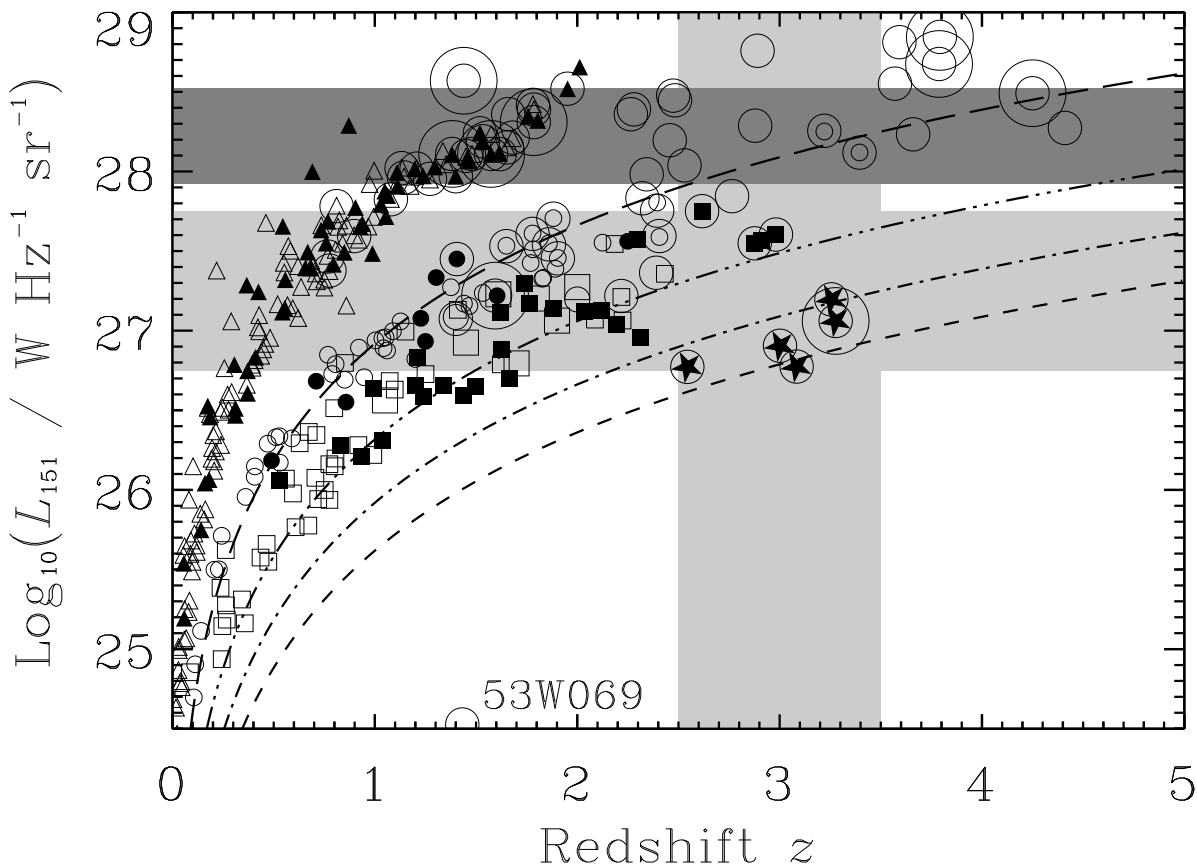}}
\end{picture}
\end{center}
{\caption[junk]{\label{fig:pz} 
The 151-MHz radio luminosity $L_{151}$ versus redshift
$z$ plane for various complete redshift surveys of radio sources,
details of which can be found in the following references: the
3CRR, revised 3C, sample (triangles) -- Laing, Riley \& Longair
(1983); the 6CE sample (circles) -- Rawlings, Eales \& Lacy (2001); 
the 7CRS, 7C Redshift Survey (squares) -- Willott et al.\ (2002b); 
five objects from the TOOT survey (stars) -- Hill \& Rawlings (2003). 
Objects classified as radio quasars, following the prescription of 
Willott et al.\ (1998), are shown as filled symbols; radio galaxies
are shown as open symbols, and these have slightly larger size in the few cases
where the redshifts are non-spectroscopic. 
The intersection of the lightly-shaded horizontal and vertical bands 
is the region of this plane [$2.5 \leq z < 3.5$ and
$26.75 \leq \log_{10} (L_{151}) < 27.75$]
central to the work of this paper. The darkly-shaded band shows the strip in
radio luminosity considered in detail by Archibald et al.\ (2001).
The dashed curves show the loci of objects
with $\alpha_{\rm rad} = 1$ and $S_{151} = 0.1 ~ \rm Jy$ (bottom curve), 
$S_{151} = 0.2 ~ \rm Jy$ (lower middle curve), $S_{151} = 0.5 ~ \rm
Jy$ (upper middle curve) and $S_{151} = 2.0 ~ \rm Jy$
(top curve). Large rings represent 
hyperluminous infrared galaxies 
associated with powerful radio sources\footnotemark, with
the smaller rings showing all those objects 
observed, but not necessarily detected, with SCUBA (Archibald et al.\ 2001; Willott 
et al.\ 2002b; this paper); note that this includes objects, like the very-low-$L_{151}$
object 53W069 (labelled), not in the complete redshift surveys. 
}}
\end{figure*}

Our SCUBA targets were chosen from two of the suite of redshift surveys of 
low-frequency-selected radio sources outlined by Rawlings (2002): namely,
(i) the {\sc TexOx}-1000, or TOOT, Survey (Hill \& Rawlings 2003); 
and (ii) parts I \& II of the 7C Redshift Survey 
(7CRS\footnotemark; Willott et al.\ 2002b).
We aimed to focus on the specific region in the 151-MHz radio luminosity 
$L_{151}$ versus redshift $z$ plane illustrated in Fig.~\ref{fig:pz}.
By studying objects in the region of intersection
of the lightly-shaded horizontal and vertical
bands of Fig.~\ref{fig:pz} we are
sensitive to the one dex range in $L_{151}$ which contributes $\sim 50$ per cent of the
radio luminosity density across the range in
cosmic time $t$ from 1.8 Gyr, at $z = 3.5$, to 2.6 Gyr at $z = 2.5$ 
(Willott et al.\ 2001b).

We used the following selection criteria: for TOOT (Hill \& Rawlings 2003), 
the TOOT00 and TOOT08 sub-regions with 
$0.1 \leq S_{151} / \rm Jy < 0.2$,
$2.5 \leq z < 3.5$ and (for a preliminary version of the survey)
$R \ltsimeq 23.5$,
yielding five targets; for 7CRS, 
$S_{151} / \rm Jy > 0.5$ and $z \geq 2.5$, yielding four potential
targets of which one, 5C6.291, was excluded from our SCUBA target list
because, with a 1.3-mm (IRAM) flux density of $65 \pm 6 ~ \rm mJy$
(Steppe et al.\ 1995), its submillimetre emission is clearly 
dominated by a bright synchrotron-emitting core.
All objects are radio quasars: for the TOOT sub-sample
this is probably because any radio galaxies meeting the $S_{151}$ and $z$ 
criteria would be too faint optically to meet the $R$ selection criterion;
for 7CRS it is because there are no spectroscopically-confirmed radio galaxies
in the target region, and none of the few objects without secure redshifts
are likely to be at suitably high redshift (Willott, Rawlings \& Blundell 2001a).
Basic data on the combined TOOT/7CRS sample can be found in
Table~\ref{tab:sample}.

\footnotetext{
Parts {\sc I \& II} 
of the 7CRS were pursued in regions of the sky covered previously
by the 5C radio surveys and, for consistency with previous papers, 
we use the 5C names throughout.
}

\footnotetext{
We deem an object to be hyperluminous if,
after making the corrections for synchrotron contamination detailed in
Table~\ref{tab:coadds}, it has a far-infrared ($30-1000 ~\mu$m) 
luminosity $L_{\rm FIR} > 10^{13} ~ \rm L_{\odot}$;
such objects have a rest-frame infrared ($1-1000 ~\mu$m) 
luminosity $L_{\rm IR} \gg 10^{13} ~ \rm L_{\odot}$
and, assuming a starburst heats the dust,
a star-formation rate $SFR \gtsimeq 1000 ~ \rm M_{\odot} 
~ yr^{-1}$ (see Fig.~\ref{fig:cosmic1}). The nine
hyperluminous objects are: 3C181, 3C298, 3C318, 3C432, 6C1045+3513
(Willott et al.\ 2002a); 4C60.07, 4C41.17, 8C1435+635 
(Archibald et al.\ 2001); TOOT08\_061 (this paper).
}

\begin{table*}
\begin{center}
\begin{tabular}{lclrrrrrr} 
\hline\hline 
 \mc{1}{l}{ Name} & \mc{1}{c}{Class}& 
 \mc{1}{c}{$z$} &
\mc{1}{c}{$\log_{10} [L_{151} /$}& \mc{1}{c}{$\log_{10} [L_{\rm core} /$}& 
\mc{1}{c}{$M_{\rm B}$}
& \mc{1}{c}{$D /$}& \mc{1}{c}{ $S_{850} /$}   &\mc{1}{c}{ $S_{450} /$} \\
 \mc{1}{l}{ } &  \mc{1}{l}{ } & 
\mc{1}{c}{}& \mc{2}{c}{W Hz$^{-1}$ sr$^{-1}$]}& 
& \mc{1}{c}{ kpc }& \mc{1}{c}{mJy} & \mc{1}{c}{mJy} \\
\hline\hline 

TOOT00\_1214 & Q(R) & 
 3.084  &   26.77  & $\approx 24.9$&  -23.0 & $\approx 38\dag$ 
& \f       2.04 $\pm$ 1.18  &\f   3.8 $\pm$ 11.1      \\
TOOT00\_1261 & Q(R) &
  2.544  &   26.77  & $\approx 24.8$& -22.8 & $\approx 24\dag$ 
& \f       1.98 $\pm$ 1.31  &\f  -18.9 $\pm$ 22.8     \\
5C6.95       &  Q(R) &
 2.877  &   27.55  &  25.6        &  -27.0  & 119.3 
& \f       2.40 $\pm$ 1.23  &\f    0.3 $\pm$ 13.4     \\
5C6.288      & Q(R) &
2.982  &    27.70 &  26.8        &  -24.6 &   7.7
& \f {\bf 16.74} $\pm$ 1.83 &\f   21.6 $\pm$ 22.7     \\
5C7.70       & Q(R) &
2.617  &    27.75 &  26.3        &  -25.1 &  14.4 
& \f       0.32  $\pm$ 1.20 &\f 5.9 $\pm$ 14.8        \\
TOOT08\_061  & Q(R) & 
3.277  &    27.06 &   -           & -27.1 & $< 41.0$  
& \f {\bf 8.59} $\pm$ 1.10 &\f {\bf 14.9}  $\pm$   5.7  \\
TOOT08\_079  & Q(S) &
3.002  &    26.91 &   -           & -23.7 & $< 42.0$ 
& \f       0.05 $\pm$ 1.14 &  \f  -2.5   $\pm$   9.1 \\
TOOT08\_094  & Q(R) & 
3.256 &   27.20  &    -           & -24.1 & $< 41.0$ 
& \f 0.91  $\pm$ 1.05 &  \f  -5.1   $\pm$   6.5 \\
 
\hline\hline
\end{tabular}
\end{center}
{\caption[Table]{\label{tab:sample} 
Basic data for the sample observed with SCUBA from
Rawlings, Hill \& Willott (2004) for the TOOT objects and
Willott et al.\ (1998) for the 7CRS objects.
Following the classification scheme of Willott et al.\ (1998), all
objects are quasars, but all show evidence for obscuration. We have therefore
sub-divided them according to our best guess as to the nature
of the obscuration: Q(R) means the quasar nucleus is seen in
lightly-reddened, transmitted light; 
Q(S) means the quasar nucleus is probably revealed by scattered, rather
than transmitted, light  -- no objects are classified Q(N), meaning a 
naked quasar nucleus. The absolute $B$ magnitudes are not corrected 
for any extinction. 
Radio core luminosities have been estimated as close as possible
to a rest-frame frequency of 5 GHz; a `-' 
denotes that existing radio data are of too poor
resolution to allow isolation of any radio core.
A $\dag$ denotes that the projected linear size $D$ is that 
between the core and an extended component on just one side
of the core. The submillimetre photometry comes from the 
programme described in Sec.~\ref{sec:submm}: detections at 
the $\geq 2\sigma$
level are given in bold type; no corrections have been made for
possible contamination by synchrotron radiation. 
}} 
\end{table*}

\subsection{SCUBA photometry and origin of the submillimetre emission}
\label{sec:submm}

The sample of TOOT/7CRS radio sources was observed in photometry mode 
with the SCUBA
bolometer array at the JCMT on the nights 2001
February 24-26, and
2001 July 30-31. Observations were made simultaneously at 850 
$\mu$m and 450 $\mu$m. Our aim was to reach an rms sensitivity of
$\sim 1$ mJy at 850 $\mu$m if no clear ($> 5 \sigma$) detection was
achieved earlier. The data were reduced using the SURF package,
following the same methods as Archibald et al.\ (2001) and 
Willott et al.\ (2002a). The resulting submillimetre
flux densities are given in Table~\ref{tab:sample}. Note that
the error bars on the photometry do not include systematic
uncertainties in the absolute flux calibration which, from inter- and
intra-night variations, we estimate to be $\approx 15$ and
$\approx 30$ per cent at $850$- and $450$-$\mu$m respectively.

\begin{figure*}
\begin{center}
\setlength{\unitlength}{1mm}
\begin{picture}(150,120)
\put(-50,-30){\includegraphics{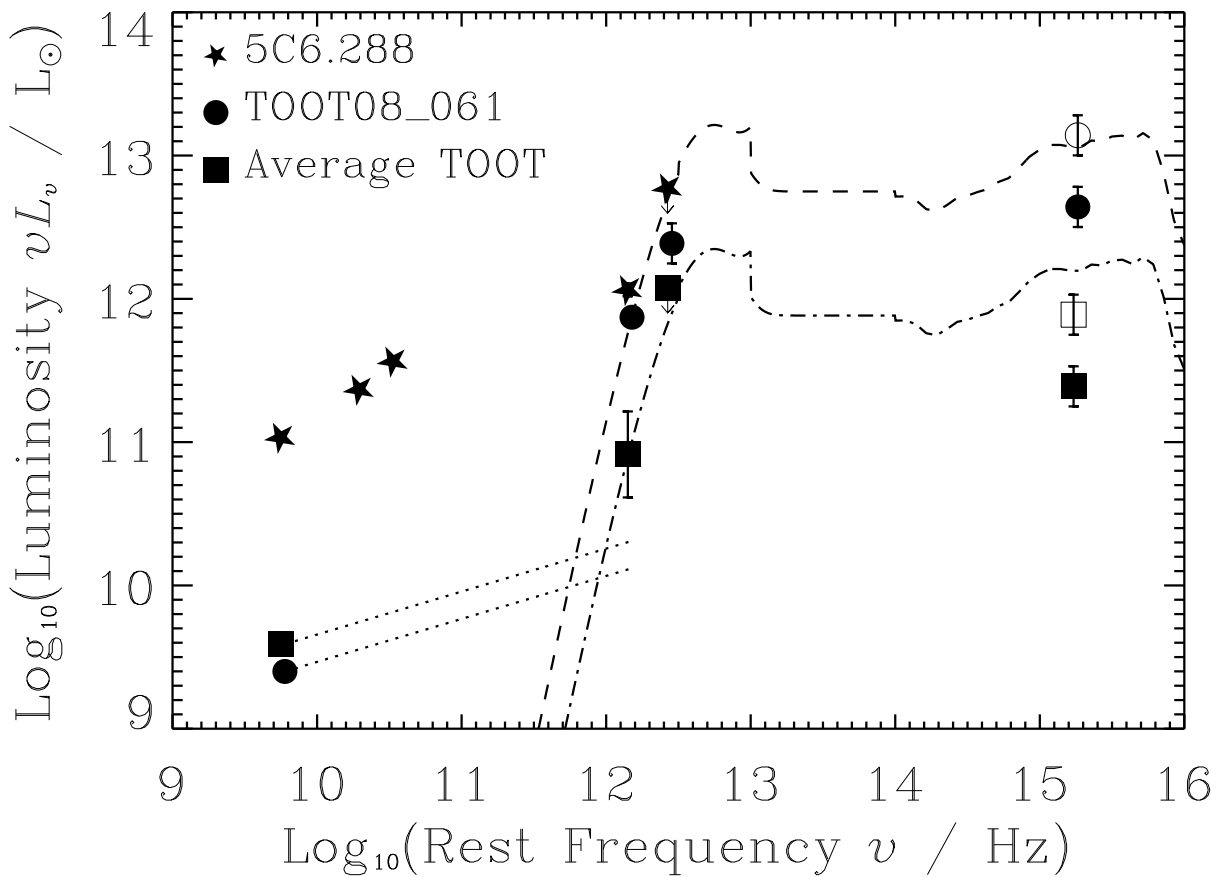}}
\end{picture}
\end{center}
{\caption[junk]{\label{fig:sed} 
The spectral energy distributions (SEDs) of objects in the
TOOT/7CRS sample. Data points are shown by the following
symbols: TOOT08\_061 (circles); variance-weighted mean of the other four
high-redshift TOOT quasars (squares, labelled `Average TOOT'); 5C6.288 (stars). 
Error bars represent just the random errors.
Open symbols represent data points corrected for dust extinction
(amounting to $A_{\rm V}=0.5$ of Milky-Way-type dust)
in the host galaxy. The dotted lines show extrapolations of the
radio flux densities assuming a spectral index of 0.7. The
dot-dashed lines show model SEDs, normalised to 
the $850 ~ \mu$m points, and 
based on the average quasar SED of Rowan-Robinson (1995)
as detailed in Sec.~\ref{sec:sed}. 
}}
\end{figure*}

Just two of the eight objects are securely detected at 850 $\mu$m, and
only one of the eight objects at 450 $\mu$m. In Fig.~\ref{fig:sed} we show the
spectral energy distributions (SEDs) of the two detected 
objects. From the comparison of these
SEDs with other well-studied objects, it seems clear that
the submillimetre emission of the one 
object detected at 450 $\mu$m (TOOT08\_061) is dominated by
dust emission. The submillimetre emission from
the other detected object (5C6.288) 
is probably dominated by synchrotron emission, although a
significant contribution from dust cannot be ruled out.

We have coadded the 850- and 450-$\mu$m flux densities of the other
six targets obtaining an $\approx 2.5 \sigma$ `statistical'
detection at 850 $\mu$m given by the variance-weighted mean 
$1.22 \pm 0.48 ~ \rm mJy$ at 850 $\mu$m, and data 
consistent with pure noise at 450 $\mu$m (corresponding to
a $2 \sigma$ upper limit $\approx 9 ~ \rm mJy$). Repeating this 
exercise for just the four TOOT objects without clear submillimetre detections
produces similar results ($1.17 \pm 0.58 ~ \rm mJy$ at 850 $\mu$m), 
although the significance of the statistical detection at 850 $\mu$m
falls to $\approx 2 \sigma$.
We conclude that more TOOT and 7CRS objects would be securely detected at 
850 $\mu$m with only slightly deeper photometric observations, although
we cannot rule out a `bimodal' distribution in 850-$\mu$m flux
density $S_{850}$ 
such that some sources would only be detected with
extremely deep exposures. 

\section{The submillimetre properties of typical high-redshift radio sources}
\label{sec:toot}

\subsection{Spectral Energy Distributions}
\label{sec:sed}

We consider first the SED of TOOT08\_061 (Fig.~\ref{fig:sed}),
the one TOOT/7CRS object to be detected at both $450$ and $850 ~\mu$m.
We used the $850 \mu$m detection to fix the normalisation of a template
quasar SED in the following manner. The
model SEDs are the sum of three components: (i) a `starburst' component modelled
by an isothermal optically-thin grey body of temperature 40 K and 
emissivity index $\beta = 2$ (see Priddey \& McMahon 2001 for 
a justification of this choice of parameters); 
(ii) a `Seyfert' component, dominant at $3-30 ~ \mu$m
from Rowan-Robinson (1995); and (iii) a quasar component, dominant
in the optical and near-IR, 
again from Rowan-Robinson (1995). The frequency-integrated luminosities of the
three components are fixed in the ratio 0.3:0.3:1 so as to reflect 
the typical values derived by Rowan-Robinson (1995). This SED is
consistent with the $450 \mu$m data point given that it is
associated with large random and systematic errors. Accounting for 
host galaxy reddening, the model SED also fits well to
the optical luminosity of TOOT08\_061. Integrating under the SED,
we estimate a bolometric luminosity $\sim 7 \times 10^{13} ~ \rm L_{\odot}$,
and since the rest-frame infrared ($1-1000 ~\mu$m) luminosity $L_{\rm IR}$
totals $\sim 4 \times 10^{13} ~ \rm L_{\odot}$, TOOT08\_061 
is the one object in the TOOT/7CRS sample with
sufficient dust emission to classify the object as a 
hyperluminous infrared galaxy  (e.g.\ Rowan-Robinson 2000). 

The correlation between 
submillimetre luminosity and bolometric luminosity for distant radio
quasars (Willott et al.\ 2002a) makes it no surprise that the
brightest submillimetre source in our TOOT/7CRS sample
is also the most optically-luminous quasar (see Table~\ref{tab:sample}).
The scatter in this correlation means that it is equally
unsurprising that the other optically-luminous quasar, 5C6.95, 
is detected only at the $\approx 2 \sigma$ level with SCUBA.

One guide we have to the more typical properties of high-redshift radio sources
comes from an `average' TOOT SED, calculated from simple
arithmetic mean values at each frequency, excluding data on 
TOOT08\_061. This SED, which is clearly biased to a somewhat lower normalisation by the
exclusion of TOOT08\_061, is also plotted in Fig.~\ref{fig:sed}. 
Synchrotron contamination of the
$S_{850} \sim 1 ~ \rm mJy$ statistical detection is a
slight concern, but, as is illustrated in Fig.~\ref{fig:sed},
extrapolation from radio to sub-mm wavelengths with
$\alpha_{\rm rad}=0.7$ [the value for the core
of 5C6.288, and seemingly, from the discussion of Willott et al.\ (2002a), 
a slightly flatter value than is typical for lobe-dominated radio
sources], does not provide a dominant contribution at $850 ~ \rm \mu$m
even if all the NVSS flux density arises from the core. 
Again fixing the normalisation of a 
standard quasar SED by the $S_{850}$ value 
(Fig.~\ref{fig:sed}), we find consistency with the 
$450 \mu$m upper limit and, by integration of the SED, an infrared luminosity 
$L_{\rm IR} \sim 5 \times 10^{12} ~ \rm L_{\odot}$, dominated presumably
by emission from dust.

Accounting for a small amount of dust extinction 
(amounting to $A_{\rm V}=0.5$ of Milky-Way-type dust as is sufficient to
explain the red colours and narrow-Ly$\alpha$ lines of the TOOT quasars; Rawlings et al.\ 2004), 
the model SED looks a little too bright at
optical wavelengths but, given the crudeness both of the
averaging process and the dust extinction correction, the difference 
between model and data
is not very significant. Thus, to the very limited
extent to which we can currently define a typical SED for 
high-redshift TOOT sources, they look similar in shape to those of the 
low-redshift quasars studied by Rowan-Robinson (1995).
The normalisation inferred for the average SED of the TOOT quasars is such 
that, in terms of infrared luminosity, typical TOOT quasars
lie comfortably within the spread exhibited by 
low-redshift, optically-selected quasars. They are naturally
classified alongside low-redshift quasars which 
have similar far-IR luminosities to ultraluminous infrared 
galaxies; about one-third of the quasars in optically-selected
low-redshift samples have these properties
(Rowan-Robinson 1995).

There is of course a continuing debate as to whether 
star formation or an AGN dominates
the heating of
the dust responsible for the far-infrared
emission from ultraluminous infrared quasars. 
We can add little to this debate, noting simply that
integration of the average TOOT SED of Fig.~\ref{fig:sed} gives a 
rest-frame far-infrared ($30-1000 ~ \mu$m) 
luminosity $L_{\rm FIR} \sim 2 \times 10^{12} ~ \rm L_{\odot}$ 
which, in the absence of any quasar heating of the cooler dust, 
corresponds to a less-than-extreme 
star-formation rate $SFR \sim 200 ~\rm M_{\odot} ~ yr^{-1}$ (e.g.\ from
equation [1] of Hughes et al.\ 1997). Given the 
possibility of quasar heating, this estimate of $SFR$
should probably be taken as an upper limit. 

\subsection{Correlations with $L_{151}$ and redshift}
\label{sec:correlation}

We have calculated variance-weighted
mean values of $S_{850}$ in various bins
of 151-MHz radio luminosity $L_{151}$ and redshift $z$ using
the 66 radio sources with SCUBA data plotted with the smaller rings in 
Fig.~\ref{fig:pz}. We eliminated from consideration the 
two sources whose 850$\mu$m data are dominated by synchrotron
emission (5C6.288 and 6C0902+34), and we made two
assumptions concerning synchrotron contamination of the other
objects: (i) zero synchrotron contamination at 850$\mu$m for all objects;
(ii) contamination at the levels detailed in the
caption of Table~\ref{tab:coadds}. These assumptions should bracket the 
uncertainties due to synchrotron contamination.
Table~\ref{tab:coadds} contains the results of this analysis,
and far-infrared luminosities $L_{\rm FIR}$
calculated from these results are plotted against redshift $z$
in Fig.~\ref{fig:cosmic1}.

\begin{table*}
\begin{center}
\begin{tabular}{lllllll} 
\hline\hline 
 \mc{1}{c}{Sample} & \mc{1}{c}{Range in $L_{151}$} & SC & 
 \mc{1}{c}{$S_{850}$ at $z=1$} & \mc{1}{c}{$S_{850}$ at $z=2$}& 
 \mc{1}{c}{$S_{850}$ at $z=3$} & \mc{1}{c}{$S_{850}$ at $z=4$}  \\

\hline\hline 

All in Fig.~\ref{fig:pz}
& $26.75 \leq L_{151} \leq 27.75$ & n & $0.85 \pm 0.43$ (5) & 
$1.87 \pm 0.35$ (9) & $2.40 \pm 0.44$ (7) & - \\
& & y & 
$0.12 \pm 0.43$ (5) & 
$1.69 \pm 0.35$ (9) &  
$2.40 \pm 0.44$ (7) & - \\

All in Fig.~\ref{fig:pz}
& $27.90 \leq L_{151} \leq 28.50$ & n & $2.52 \pm 0.36$ (8) & 
$2.97 \pm 0.26$ (16) & $2.63 \pm 0.52$ (3) & $5.84 \pm 0.59$ (3) \\
& & y &
$1.55 \pm 0.36$ (8) & 
$1.82 \pm 0.26$ (16) & $2.47 \pm 0.52$ (3) & 
$5.77 \pm 0.59$ (3) \\

Archibald et al.\ & $27.90 \leq L_{151} \leq 28.50$ & n & $1.80 \pm 0.41$ (6) 
&  
$2.10 \pm 0.30$ (11) & $2.63 \pm 0.52$ (3) & $5.84 \pm 0.59$ (3) \\
& & y & $0.98 \pm 0.41$ (6)
& $0.97 \pm 0.30$ (11) & $2.47 \pm 0.52$ (3) & $5.77 \pm 0.59$ (3) \\

Archibald et al.\ & All $L_{151}$ & n & $0.96 \pm 0.27$ (13) & 
$1.60 \pm 0.22$ (21) & $2.82 \pm 0.41$ (5) & $7.38 \pm 0.38$ (7) \\
& & y & $0.20 \pm 0.27$ (13)
& $0.89 \pm 0.22$ (21) & $2.70 \pm 0.41$ (5) & $7.32 \pm 0.38$ (7) \\

\hline\hline
\end{tabular}
\end{center}
{\caption[Table]{\label{tab:coadds} 
Variance-weighted mean values of 850-$\mu$m flux density $S_{850}$
for samples of radio sources covering
various ranges of 151-MHz luminosity $L_{151}$ and redshift $z$.
These are calculated for both all the SCUBA-observed radio sources
shown in Fig.~\ref{fig:pz} of this paper, and for just those objects in the
Archibald et al.\ (2001) study. Synchrotron-dominated
objects (5C6.288 and 6C0902+34)
have been excluded from the analysis, and values calculated
both with (SC `y') and without (SC `n')
synchrotron correction are given. 
The synchrotron-corrected values are taken
from Archibald et al.\, excepting for the objects from Willott et al.\
(2002a) and this paper. For the Willott et al.\ objects, we
assumed a radio-to-submillimetre spectral index $\alpha = 0.7$,
together with other constraints discussed by Willott et al., to estimate
the following synchrotron contaminations, expressed as 
fractions of the observed $S_{850}$: 3C181 (0.06); 3C191 (0.30);
3C205 (0.52); 3C268.4 (0.58); 3C270.1 (0.40); 3C280.1 (0.35);
3C298 (0.38); 3C318 (0.00); 3C432 (0.05);
6C0955+3844 (0.00); 6C1045+3513 (0.05). 
For the TOOT and 7C objects
we estimate zero synchrotron contamination because 
such low-$S_{151}$ objects can only contribute significantly at 850 $\mu$m
if their radio flux is concentrated in a compact core
(see Fig.\ 2).
The redshift bins used were 
$0.5-1.5$, $1.5-2.5$, $2.5-3.5$ and $3.5-4.5$. The number of
objects involved in calculating
each average are given in brackets.
}} 
\end{table*}

\subsubsection{Correlation with $L_{151}$}
\label{sec:correlationl151}

\begin{figure*}
\begin{center}
\setlength{\unitlength}{1mm}
\begin{picture}(150,120)
\put(-50,-30){\includegraphics{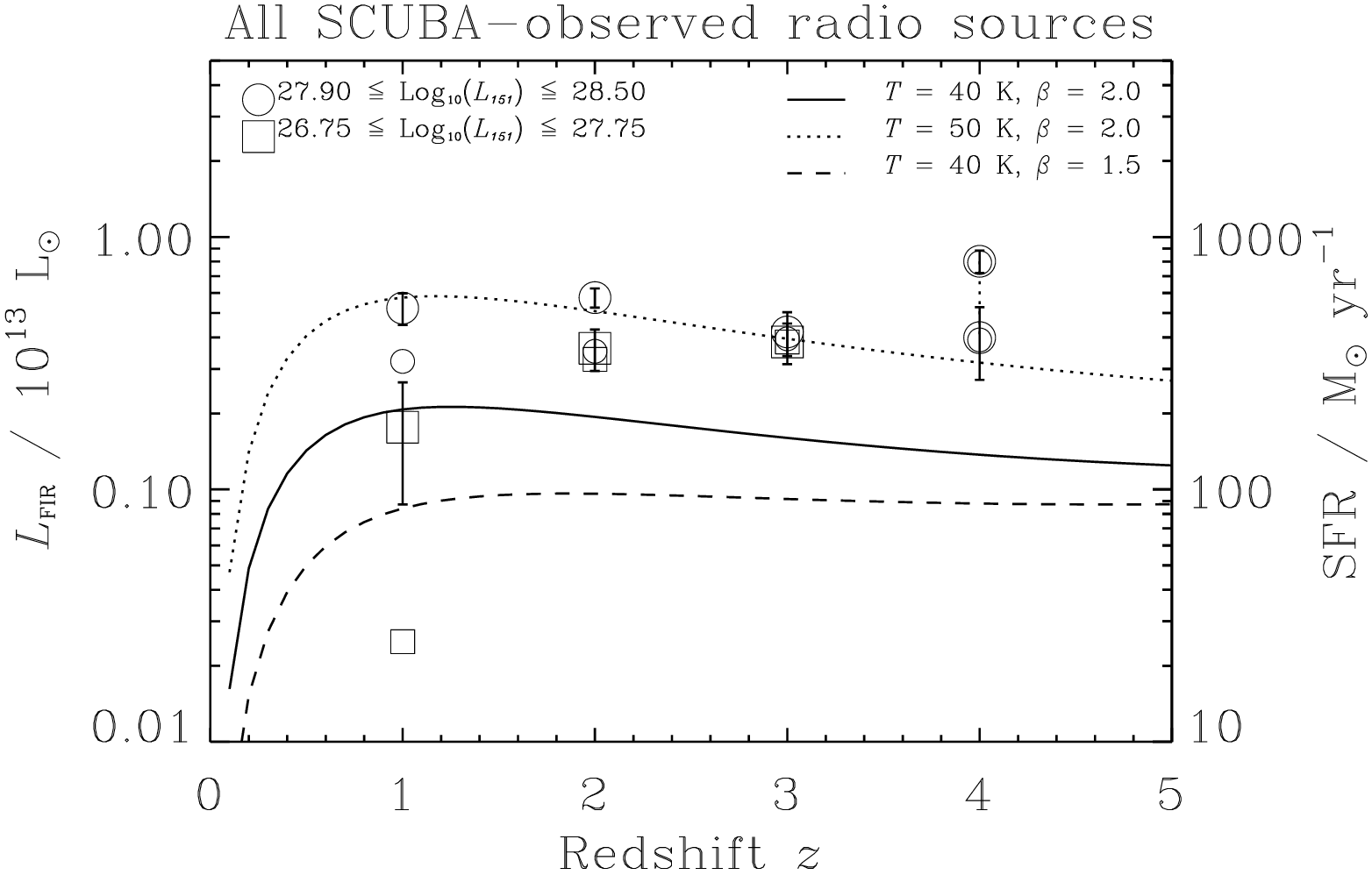}}
\end{picture}
\end{center}
{\caption[junk]{\label{fig:cosmic1} 
The variance-weighted mean
far-infrared ($30-1000 ~\mu$m) luminosity $L_{\rm FIR}$
of radio sources in selected ranges of 151-MHz luminosity $L_{151}$
versus redshift $z$
(in bins $0.5-1.5$, $1.5-2.5$, $2.5-3.5$ and $3.5-4.5$).
All objects identified as observed with SCUBA in Fig.~\ref{fig:pz},
barring three synchrotron-dominated objects, were used to calculate these
averages (see Table~\ref{tab:coadds}). 
We have calculated $L_{\rm FIR}$ using an isothermal 
optically-thin grey body spectrum characterised by a temperature $T=40 ~ K$
and emissivity index $\beta=2$; the locus of an $S_{850}=1 ~ \rm mJy$
source with this spectrum is shown by the solid line, 
with the effects of varying $T$ and $\beta$ being shown by the 
dotted line ($T=50$, $\beta=2$) and the dashed line ($T=40$, $\beta=1.5$).
$L_{\rm FIR}$ is converted to star formation rate 
$SFR$ using eqn. (1) of Hughes et al.\ (1997), adopting $\epsilon = 1$.
The large open squares
represent radio sources with $26.75 \leq \log_{10} (L_{151}) < 27.75$.
The large open circles
represent radio sources with $27.9 \leq \log_{10} (L_{151}) < 28.5$,
i.e.\ the high-luminosity strip considered by Archibald et al.\ (2001); two
different values are plotted at $z=4$, the lower point representing just the 
two 6C* objects, the upper point representing all three objects in this bin.
The smaller symbols, with error bars not plotted, show the mean
values after correction
for synchrotron contamination (see Table~\ref{tab:coadds}).
Note that this plot is affected by the various potential systematic biases 
discussed in Sec.~\ref{sec:toot}. 
}}
\end{figure*}

The baseline in $L_{151}$ opened up at high redshift by the TOOT/7CRS
objects allows us to look directly for a correlation
between $L_{\rm FIR}$ and $L_{151}$ at early cosmic epochs. The
results of a straight comparison of the eight TOOT/7CRS objects
with the six more radio luminous but similar-redshift
objects (8C1039+68, MG1016+058, 4C24.28, 4C28.58, 6C0902+34 and 
6C1232+39) from the Archibald et al.\ (2001) study, excluding the two 
synchrotron-dominated objects (5C6.288 and 6C0902+34), are
included in Table~\ref{tab:coadds} and 
Fig.~\ref{fig:cosmic1}. Over this restricted redshift range
($2.5 \leq z < 3.5$)
we see that, despite a one dex difference in average $L_{151}$, there
is no difference in average $S_{850}$. How does this 
constrain any correlation between $L_{151}$ and $L_{\rm FIR}$?

First, we neglect synchrotron contamination and consider 
the effects of the submillimetre difference between radio quasars and radio galaxies
discovered by Willott et al.\ (2002a).
The average value of $S_{850}$ for the $z \sim 3$ TOOT/7CRS  
objects is likely to be biased high by the exclusion of 
radio galaxies (see Sec.~\ref{sec:selection}), and the average $S_{850}$ for 
the higher-$L_{151}$ $z \sim 3$ objects biased low by the exclusion of 
radio quasars (Archibald et al.\ 2001). We can estimate the
relevant bias factors as follows. As the higher-$L_{151}$ sources have similar
values of $L_{151}$ to the radio quasars studied by Willott et al.\ (2002a), 
it is reasonable to assume that they have a similar quasar fraction 
($\approx 0.4$; Willott et al. 2000) and a similar submillimetre 
difference (radio quasars being, after correction for
synchrotron contamination, a factor $\sim 3$ brighter than the 
radio galaxies at 850 $\mu$m), so that the average $S_{850}$ would be expected to be
a factor $\sim 2$ higher in a genuinely complete sample.
For a genuinely complete TOOT/7CRS sample, the average $S_{850}$
is likely to be lower, but by a smaller factor because the quasar fraction is likely to be 
similar (Willott et al. 2000) and there will a lower contribution to the average by the
excluded objects. Overall, therefore, for $z \sim 3$ complete samples, the high-$L_{151}$ 
point in Fig.~\ref{fig:cosmic1} 
is expected to be higher than the low-$L_{151}$ point by a 
factor $\sim 2$. This is a similar separation to that seen
between the high- and low-$L_{151}$ points at $z \sim 1$ and $z \sim 2$
in Fig.~\ref{fig:cosmic1}, and these do not need large corrections
for object exclusion because the Willott et al.\ (2002a) programme 
ensured that the SCUBA-targeted radio quasars and radio galaxies ended up
in roughly the same ratio as the quasar fraction. 

We conclude that, although
significant only at the $\sim 2 \sigma$ level (see also
Sec.~\ref{sec:correlationz}), there is evidence for a $L_{\rm FIR}$--$L_{151}$
correlation in Fig.~\ref{fig:cosmic1}. An increase in $L_{\rm FIR}$ by a factor
$\sim 2$ as $L_{151}$ increases by a factor $\sim 10$ are consistent
with a model in which $L_{\rm FIR} \propto L_{151}^{p}$, provided
$p \ltsimeq 0.5$. 

If we adopt the synchrotron-corrected values from Table~\ref{tab:coadds},
then a somewhat different picture emerges (Fig~\ref{fig:cosmic1}). Most 
noticeably, there now exists a large (factor $\sim 10$) difference between the high- and
low-$L_{151}$ objects at $z \sim 1$. Synchrotron corrections have 
most effect at low redshift because 850$\mu$m corresponds to a lower
rest-frame frequency, and most of the SCUBA-targeted galaxies are 
from the radio-bright 3C survey\footnotemark. This might be interpreted as
evidence that we have, if anything, underestimated the strength and slope of the 
$L_{\rm FIR}$--$L_{151}$ correlation, or alternatively that the (uncertain) synchrotron
corrections have been too severe. However, it is also not completely implausible
that the synchrotron corrections are not severe enough 
[although see Willott et al.\ (2002a) for arguments to the contrary]. If this were the case, 
most 850-$\mu$m detections at $z \sim 1$ could be dominated by synchrotron contamination
and the difference in dust submillimetre luminosity between radio quasars
and radio galaxies inferred by Willott et al.\ (2002a) might have to be reduced. The
correct interpretation of Fig.~\ref{fig:cosmic1} might then be that there is little evidence for
a $L_{\rm FIR}$-$L_{151}$ correlation at any redshift, but there is 
a dramatic drop in $L_{\rm FIR}$ at $z \sim 1$ (see Fig~\ref{fig:cosmic1}); we return to this
point in Sec.~\ref{sec:correlationz}.

\footnotetext{
Note that the method of synchrotron 
correction adopted by Archibald et al.\ (2001)
tends to set small positive values of $S_{850}$ of low-$z$, typically 3C, sources to zero, 
whilst leaving negative values unchanged.
This may result in `corrected' values which are biased low. 
}

\subsubsection{Correlation with redshift}
\label{sec:correlationz}

\begin{figure*}
\begin{center}
\setlength{\unitlength}{1mm}
\begin{picture}(150,120)
\put(-50,-30){\includegraphics{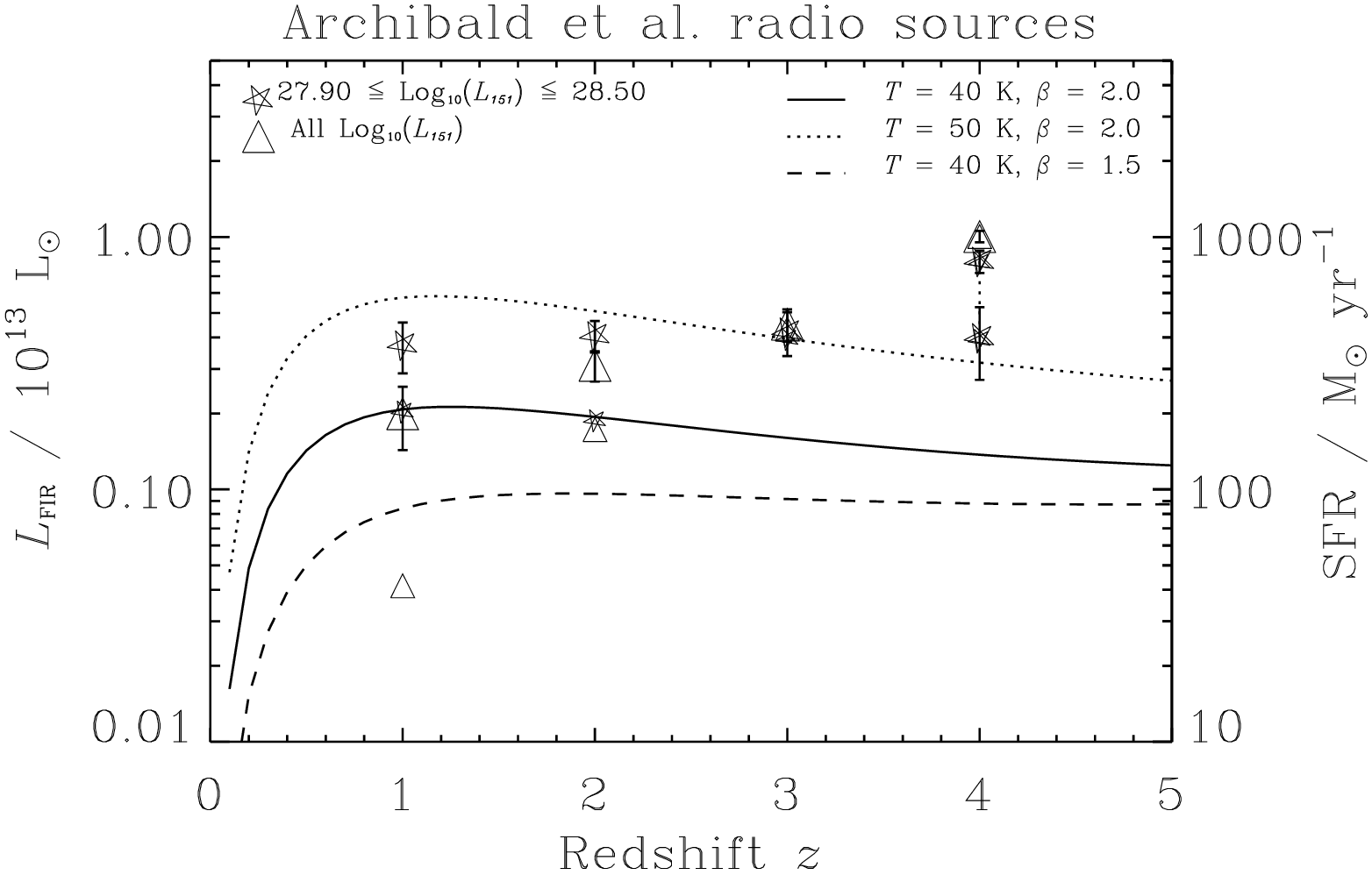}}
\end{picture}
\end{center}
{\caption[junk]{\label{fig:cosmic2} 
The variance-weighted mean
far-infrared ($30-1000 ~\mu$m) luminosity $L_{\rm FIR}$
of radio sources in selected ranges of 151-MHz luminosity $L_{151}$
versus redshift $z$
(in bins $0.5-1.5$, $1.5-2.5$, $2.5-3.5$ and $3.5-4.5$). Only
objects included in the Archibald et al.\ (2001) study were used to
calculate these averages (see Table~\ref{tab:coadds}). 
The axes and lines have the same meaning as in 
Fig.~\ref{fig:cosmic1}.
The large stars represent just the Archibald et al.\ (2001) objects
in the $27.9 \leq \log_{10} (L_{151}) < 28.5$ strip (two
values are plotted at $z=4$, the lower point representing just the 
two 6C* objects, the upper point representing all three objects); 
the large 
triangles represent all the Archibald et al.\ (2001) objects, irrespective
of radio luminosity. 
The smaller symbols, with error bars not plotted, show the
values after correction 
for synchrotron contamination (see Table~\ref{tab:coadds}). 
Note that this plot is affected by the 
various potential systematic biases discussed in
Sec.~\ref{sec:toot}.
}}
\end{figure*}

The similarity between high-$z$ TOOT/7CRS quasars and some
low-redshift optically-selected quasars, 
discussed in Sec.~\ref{sec:sed}, suggests
we investigate whether there is any direct 
evidence for the cosmic evolution in the rest-frame far-infrared
luminosities $L_{\rm FIR}$ of radio sources of `typical' radio luminosity,
i.e.\ across the lightly-shaded horizontal strip in 
Fig.~\ref{fig:pz}. The results of this investigation are 
included in Table~\ref{tab:coadds} and Fig.~\ref{fig:cosmic1}.

Again, we start our discussion by adopting the assumption of 
negligible synchrotron contamination.
The evidence for any systematic rise in
$L_{\rm FIR}$ with redshift is then weak ($< 2 \sigma$)
and, as was the case for the hints of a correlation between $L_{\rm FIR}$
and $L_{151}$ (Sec.~\ref{sec:correlationl151}),
driven by the low average $L_{\rm FIR}$ of the low-$L_{151}$, 
$z \sim 1$ bin. Mindful of the many biases discussed in
Sec.~\ref{sec:correlationl151},  
we conclude that, neglecting corrections for
synchrotron contamination, there is no firm evidence
for a cosmic rise in $L_{\rm FIR}$ for typical radio luminosity sources.
As we have already noted in Sec.~\ref{sec:correlationl151},
adopting the synchrotron-corrected values from Table~\ref{tab:coadds} means that 
there is much stronger evidence for cosmic evolution in the sense
of a dramatic drop in $L_{\rm FIR}$ at $z \sim 1$.

Evidence for any cosmic rise in $L_{\rm FIR}$ for the 
high-$L_{151}$ sources (Fig.~\ref{fig:cosmic1}),
with our without synchrotron correction, is very weak, driven,
this time, by a high average $L_{\rm FIR}$ in the $z \sim 4$ bin, a
redshift range not yet probed by SCUBA observations of low-$L_{151}$,
i.e.\ TOOT, sources.
There are just three objects
(6C*0032+412, 8C1435+635, 6C*0140+326) in this bin, 
all from the Archibald et al.\ (2001) SCUBA survey, 
so small number statistics and selection biases are 
extremely important. These objects are from
`filtered', i.e.\ steep-radio-spectrum-selected, rather than
complete-flux-density-limited, samples (8C, Lacy et al.\ 1994;
6C*, Jarvis et al.\ 2001a). For such
samples, 6C* is the only one in which
it has been demonstrated that potentially 
high-$z$ objects have {\it not} been missed because of low optical emission 
line strengths (Jarvis et al.\ 2001a). 
Given that scaling relations between luminosities in different 
wavebands are normally positive, the use of
samples in which low-emission-line-strength objects have been  
missed would have the effect of
biasing the average $S_{850}$ value high. Considering just the two 6C*
objects, we see in Fig.~\ref{fig:cosmic1}
no evidence that $L_{\rm FIR}$ is any higher at 
$z \sim 4$ than at lower redshift. A rise in average $L_{\rm FIR}$
only emerges if one adds in the hyperluminous 
object 8C1435+635. We caution against putting 
much weight on either this rise, and its lower formal error bar, 
because the parent (8C) sample does not have a
well-defined redshift completeness. It should also be
noticed that any $z \sim 4$ radio quasars are likely to have been missed 
by all filtered samples, including 6C*, because their prominent 
radio cores can mask steep-spectrum extended emission.
The inclusion of such objects would tend to increase the average value 
of $S_{850}$ (Willott et al.\ 2002a).

Adopting the assumption of negligible synchrotron contamination,
the weakness of any correlations with redshift in
Fig.~\ref{fig:cosmic1} means we should re-examine the
seemingly strong evidence for a cosmic rise in $L_{\rm FIR}$ 
from the study of Archibald et al.\ (2001). Binned data
from their study are presented in
Table~\ref{tab:coadds} and plotted in Fig.~\ref{fig:cosmic2}.
Considering just those objects in the high-$L_{151}$ strip of Fig.~\ref{fig:pz}
we see that, as in Fig.~\ref{fig:cosmic1}, evidence for 
a cosmic rise comes from just the $z \sim 4$ bin, and
is subject to the same worries concerning small-number
statistics and selection biases. It is only by averaging all
the Archibald et al.\ data, regardless of $L_{151}$, that 
a seemingly monotonic rise in $L_{\rm FIR}$ with $z$ emerges
(Fig.~\ref{fig:cosmic2}). There are several potential problems  
with interpreting this correctly. First, there is a strong and systematic increase in the
average $L_{151}$ of objects from the lowest-redshift bin
(mean $L_{151} = 27.5$) to the highest-redshift bin (mean $L_{151} = 28.6$):
this could account for nearly all of the `cosmic' rise in $L_{\rm FIR}$ given
a correlation of the form $L_{\rm FIR} \propto L_{151}^{p}$, with $p \ltsimeq 0.5$
as discussed in Sec.~\ref{sec:correlationl151}.
Second, the average value of the $z \sim 4$ bin is, 
despite the low formal error bar, calculated from objects drawn from
various filtered surveys, and therefore subject to the possible
selection biases discussed in Sec.~\ref{sec:correlationl151}.
Third, the reality and magnitude of any cosmic rise 
in $L_{\rm FIR}$ is strongly dependent on the assumed dust template.
Our choice of dust template
(dust temperature $T=40 ~ \rm K$, emissivity index $\beta = 2.0$)
is motivated by fits to the SEDs of distant quasars by
Priddey \& McMahon (2001). For this template,
objects of a given $L_{\rm FIR}$
have higher $S_{850}$ at $z \sim 4$ than at $z \sim 1$,
whereas, as illustrated in Figs.~\ref{fig:cosmic1} and
~\ref{fig:cosmic2}, the Archibald et al.\ choice of
dust template ($T=40 ~ \rm K$, $\beta = 1.5$) means that at
a given $L_{\rm FIR}$, $S_{850}$ is virtually independent of 
redshift. 
Thus, for a given cosmology and a given dataset,
our choice of dust template will always produce less
cosmic evolution than the template adopted
by Archibald et al.\ (2001). 

It is also worth re-emphasising that the Archibald et al.\ (2001) 
study excluded radio quasars. It could be argued that this
reduces, or even eliminates, the need for the complicated corrections
for biases discussed in the context of Fig.~\ref{fig:cosmic1}
in Sec.~\ref{sec:correlationl151}. However, radio galaxies will only
show the same relationship between $L_{\rm FIR}$ and redshift as the
total (radio galaxy plus radio quasar) population if 
the probability of a radio source
being classified as radio galaxy is independent of redshift. This is
an as yet unproven assumption, and one that does not 
sit easily with the results of Hirst et al.\ (2003) and
Willott et al.\ (2002a) which suggest that both the quasar fraction and
submillimetre luminosity are functions of radio source age
which, in turn, is likely to correlate negatively with
redshift (e.g.\ Blundell \& Rawlings 1999).

\section{Concluding Remarks}
\label{sec:conclusions}

We have seen in Sec.~\ref{sec:correlationz} that
evidence for evolution of the dust properties of radio sources with redshift 
emerges only
with the adoption of the synchrotron-corrected values for $S_{850}$,
and is confused by the influence of the correlation between 
$L_{\rm FIR}$ and $L_{151}$ (Sec.~\ref{sec:correlationl151}).
Since these corrections and this correlation are very uncertain,
we conclude that, for radio sources, direct observational evidence 
that $L_{\rm FIR}$ rises systematically with redshift
is not yet compelling. Although it would be
rather surprising if an increase in the dust and gas masses of
active galaxies with redshift were not an important part of the 
overall picture (Archibald et al.\ 2001), the results of Sec.~\ref{sec:correlation}
are now in reasonable accord with the results obtained on radio-quiet 
quasars. Priddey et al.\ (2003) noted that variations in the submillimetre
detectability of optically-selected quasars with redshift are consistent 
with the K-correction of a template spectrum which does not evolve in 
normalisation with redshift. Although, submillimetre studies of radio-quiet
quasars do not suffer from the problems of synchrotron contamination, it is
interesting to note that, for radio-quiet quasars,
searches for systematic changes in dust emission with redshift 
are confused by possible correlations between $L_{\rm FIR}$
and optical luminosity (Willott, Rawlings \& Grimes 2003; Priddey et al.\ 2003).

\begin{figure*}
\begin{center}
\setlength{\unitlength}{1mm}
\begin{picture}(150,150)
\put(-50,-20){\includegraphics{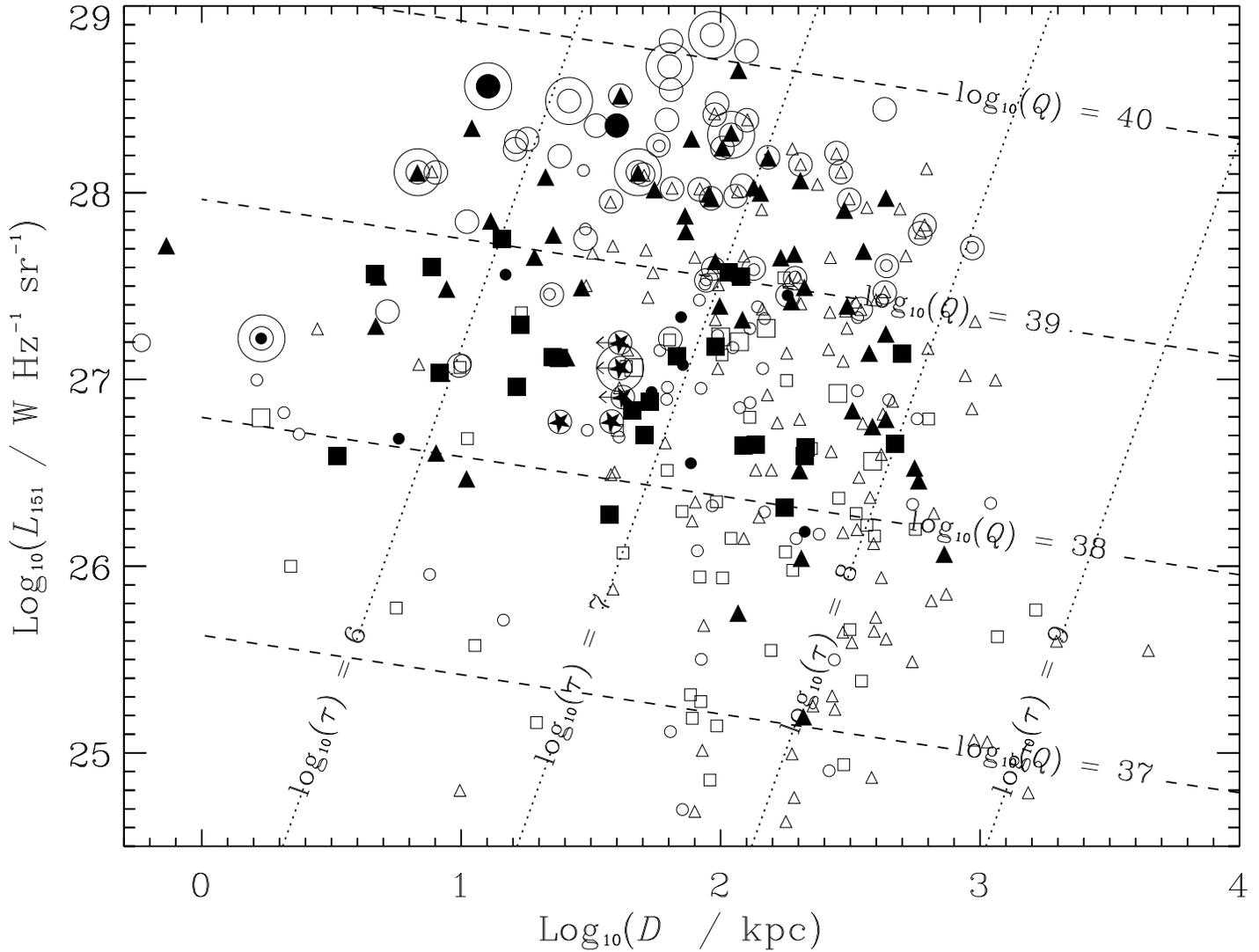}}
\end{picture}
\end{center}
{\caption[junk]{\label{fig:pd}
The 151-MHz radio luminosity $L_{151}$ versus projected linear size
$D$ plane for various complete redshift surveys of radio sources
(surveys and symbols as in Fig.~\ref{fig:pz}; filled symbols
are radio quasars, open symbols are radio galaxies). As in
Fig.~\ref{fig:pz}, large rings represent 
confirmed hyperluminous infrared galaxies and
smaller rings represent objects 
observed with SCUBA, including objects not in the
complete redshift surveys.
The lines are calculated from the radio source model of
Willott et al.\ (1999) assuming $\alpha_{\rm rad}=1$: 
the dotted lines are contours of constant 
$\log_{10}(\tau / \rm yr)$, 
where $\tau$ is the time since the
jet-triggering event, and the dashed lines are contours of 
constant $\log_{10}(Q / \rm W)$, where $Q$ is the power
in both jets. The values adopted for the fixed parameters 
of the radio source model (see Willott et al.\ 1999 for 
definitions) are as follows:  
$\theta=60^{\circ}$; $R_{T}=5$; $f=20$; $k=0$; 
$\phi = 90^{\circ}$; $\eta = 1$; $\beta=1.5$
(here $\beta$ is the slope of the density power law, not the
dust emissivity index); $c_{1}=2.3$;
$n_{100} = 3000 ~ \rm e^{-} ~ m^{-3}$;
none of these parameters, most crucially those of the 
environment ($\beta$ and $n_{100}$), are assumed to vary with
cosmic time, and some, most notably $f$, are sufficiently 
uncertain that values of $Q$ and $\tau$ should be regarded as
order-of-magnitude estimates.

}}

\end{figure*}

In Sec.~\ref{sec:correlation} we have seen that 
hyperluminous infrared objects contribute substantially to the average values of 
$S_{850}$ used in the correlation studies.
Looking at the location in Fig.~\ref{fig:pd} of the radio-loud objects with
associated hyperluminous infrared galaxies hints at complicated physical processes 
behind any correlations.
As has been shown previously by Willott et al.\ (2002a), and was
conjectured by Blundell \& Rawlings (1999), there seems
to be a clear association of hyperluminous infrared galaxies 
with the radio sources of smallest projected linear size $D$. The underlying
physics here (see caption of Fig.~\ref{fig:pd} for details) may be that
a short time $\tau \ltsimeq 10^{7} ~ \rm yr$ since the jet-triggering event is a necessary, 
if not sufficient, condition for far-infrared hyperluminosity as at later times
dust may have been destroyed or dispersed by the expanding radio 
source (De Young 1998). With two
important exceptions, the hyperluminous objects appear to be concentrated amongst the most 
radio-luminous sources. The underlying physics here may be that 
extreme radio luminosity is strongly linked to
the presence of an extreme-luminosity quasar nucleus which in turn promotes, 
perhaps by direct heating of dust, infrared hyperluminosity. The two hyperluminous
objects at relatively low $L_{151}$ (6C1045+3513 and TOOT08\_061) 
both show, unusually for radio-loud objects,
evidence for broad absorption lines (Willott et al.\ 2002a; Rawlings et al., 2004). 
The underlying physics here may be that young bursts of less-than-extreme radio luminosity, probably
associated with the early evolution of less powerful jets
(e.g.\ Kuncic 1999), as well as BAL features may be indicative of processes triggered by
galaxy mergers (e.g.\ Canalizo \& Stockton 2001).

\section*{Acknowledgements}

It is our great pleasure to thank Kate Brand and Jonathan Rawlings
for their help with the SCUBA observations. Thanks
also to Steve Croft, Pamela Gay, Julia Riley, John
Swinbank and Joe Tufts for their contributions to the TOOT Survey,
and Katherine Blundell and Mark Lacy for their 
contributions to the 7C Redshift Survey. We also thank the anonymous
referee for a very useful report. SR thanks the PPARC for a Senior Research Fellowship.
The JCMT is operated by the Joint Astronomy Centre, on behalf of
PPARC, the Netherlands Organisation for Pure Research, and the
National Research Council of Canada.
This research has made use of the NASA/IPAC Extragalactic Database, which
is operated by the Jet Propulsion Laboratory, Caltech, under contract
with the National Aeronautics and Space Administration.

\end{document}